# Causal State Communication

Chiranjib Choudhuri, Young-Han Kim, and Urbashi Mitra

**Abstract**

The problem of state communication over a discrete memoryless channel with discrete memoryless state is studied when the state information is available strictly causally at the encoder. It is shown that block Markov encoding, in which the encoder communicates a description of the state sequence in the previous block by incorporating side information about the state sequence at the decoder, yields the minimum state estimation error. When the same channel is used to send additional independent information at the expense of a higher channel state estimation error, the optimal tradeoff between the rate of the independent information and the state estimation error is characterized via the capacity–distortion function. It is shown that any optimal tradeoff pair can be achieved via rate-splitting. These coding theorems are then extended optimally to the case of causal channel state information at the encoder using the Shannon strategy.

## I. INTRODUCTION

The problem of information transmission over channels with state (also referred to as state-dependent channels) is classical. One of the most interesting models is the scenario in which the channel state is available at the encoder either causally or noncausally. This framework has been extensively studied for independent and identically distributed (i.i.d.) states, starting from the pioneering work of Shannon [26], Kusnetov and Tsybakov [18], Gelfand and Pinsker [12], and Heegard and El Gamal [14]; see a recent survey by Keshet, Steinberg, and Merhav [16].

Most of the existing literature has focused on determining the channel capacity or devising practical capacity-achieving coding techniques for this channel. In certain communication scenarios, however, the

This research has been supported in part by the National Science Foundation under Grants CCF-0747111, CNS-0832186, CNS-0821750 (MRI), CCF-0917343, and the Office of Naval Research under Grant N00014-09-1-0700.

Chiranjib Choudhuri (cchoudhu@usc.edu) and Urbashi Mitra (ubli@usc.edu) are with the Ming Hsieh Department of Electrical Engineering, University of Southern California, University Park, Los Angeles, CA 90089, USA. Young-Han Kim (yhk@ucsd.edu) is with the Department of Electrical and Computer Engineering, University of California, San Diego, La Jolla, CA 92093, USA.

The material in this paper was presented in part in [3] and [4].

encoder may instead wish to help reveal the channel state to the decoder. In this paper, we study this problem of state communication over a discrete memoryless channel (DMC) with discrete memoryless (DM) state, in which the encoder has either strictly causal or causal state information and wishes to help reveal it to the decoder with some fidelity criterion. This problem is motivated by a wide array of applications, including multimedia information hiding in Moulin and O'Sullivan [22], digital watermarking in Chen and Wornell [2], data storage over memory with defects in Kusnetsov and Tsybakov [18] and Heegard and El Gamal [14], secret communication systems in Lee and Xiang [19], dynamic spectrum access systems in Mitola [21] and later in Devroye, Mitran, and Tarokh [10], and underwater acoustic/sonar applications in Stojanovic [27]. Each of these problems can be expressed as a problem of conveying the channel state to the decoder. For instance, the encoder may be able to monitor the interference level in the channel; it only attempts to carry out communication when the interference level is low and additionally assists the decoder in estimating the interference for better decoder performance. We show that block Markov encoding, in which the encoder communicates a description of the state sequence in the previous block by incorporating side information about the state sequence at the decoder, is optimal for communicating the state when the state information is *strictly causally* available at the encoder. For the causal case, this block Markov coding scheme coupled with incorporating the current channel state using the Shannon strategy turns out to be optimal.

This same channel can also be used to send additional independent information. This is, however, accomplished at the expense of a higher channel state estimation error. We characterize the tradeoff between the amount of independent information that can be reliably transmitted and the accuracy at which the decoder can estimate the channel state via the *capacity–distortion function*, which is to be distinguished from the usual rate–distortion function in source coding. We show that any optimal tradeoff can be achieved via rate-splitting, whereby the encoder appropriately allocates its rate between information transmission and state communication.

The problem of joint communication and state estimation was introduced in [29], which studied the capacity–distortion tradeoff for the Gaussian channel with additive Gaussian state when the state information is *noncausally* available at the encoder; see Sutivong [28] for the general case. The other extreme case was studied later in [31], in which both the encoder and the decoder are assumed to be *oblivious* of the channel state; the capacity of the channel subject to a distortion constraint is determined. This paper connects these two sets of prior results by considering causal (i.e., temporally partial) information of the state at the encoder.

Note that the problem of communicating the causally (or noncausally) available state and independent information over a state-dependent channel was also studied in [17] and its dual problem of communicating independent information while masking the state was studied by Merhav and Shamai [20]. Instead of reconstructing the state in some fidelity criterion, however, the focus in [17] was the optimal tradeoff between the information transmission rate and the state uncertainty reduction rate (the list decoding exponent of the state). We will later elucidate the connection between the results in [17] and our results.

The rest of this paper is organized as follows. Section II describes the basic channel model with discrete alphabets, characterizes the minimum distortion in estimating the state, establishes its achievability and proves the converse part of the theorem. Section III extends the results to the information rate–distortion tradeoff setting, wherein we evaluate the capacity–distortion function with strictly causal state at the encoder. Since the intuition gained from the study of the strictly causal setup carries over when the encoder has causal knowledge of the state sequence, the causal case is treated only briefly in Section IV with key examples provided for the causal case. Finally, Section V concludes the paper.

Throughout the paper, we closely follow the notation in [11]. In particular, a random variable is denoted by an upper case letter (e.g., $X, Y, Z$) and its realization is denoted by a lower case letter (e.g., $x, y, z$). The shorthand notation $X^n$ is used to denote the tuple (or the column vector) of random variables $(X_1, \ldots, X_n)$, and $x^n$ is used to denote their realizations. The notation $X^n \sim p(x^n)$ means that $p(x^n)$ is the probability mass function (pmf) of the random vector $X^n$. Similarly, $Y^n | \{X^n = x^n\} \sim p(y^n|x^n)$ means that $p(y^n|x^n)$ is the conditional pmf of $Y^n$ given $\{X^n = x^n\}$. For $X \sim p(x)$ and $\epsilon \in (0,1)$, we define the set of $\epsilon$-typical $n$-sequences $x^n$ (or the typical set in short) [24] as

$$\mathcal{T}_\epsilon^{(n)}(X) = \left\{ x^n \colon \left| |\{i \colon x_i = x\}|/n - p(x) \right| \leq \epsilon p(x) \text{ for all } x \in \mathcal{X} \right\}.$$

We say that $X \to Y \to Z$ form a Markov chain if $p(x,y,z) = p(x)p(y|x)p(z|y)$, that is, $X$ and $Z$ are conditionally independent of each other given $Y$. Finally, $\mathsf{C}(x) = (1/2)\log(1+x)$ denotes the Gaussian capacity function.

## II. PROBLEM SETUP AND MAIN RESULT

Consider a point-to-point communication system with state depicted in Fig. 1. Suppose that the encoder has *strictly causal* access to the channel state sequence $S^n$ and wishes to communicate the state to the decoder. We assume a DMC with a DM state model $(\mathcal{X} \times \mathcal{S}, p(y|x,s)p(s), \mathcal{Y})$ that consists of a finite input alphabet $\mathcal{X}$, a finite output alphabet $\mathcal{Y}$, a finite state alphabet $\mathcal{S}$, and a collection of conditional pmfs $p(y|x,s)$ on $\mathcal{Y}$. The *channel* is memoryless in the sense that, without feedback, $p(y^n|x^n, s^n) = \prod_{i=1}^n p_{Y|X,S}(y_i|x_i, s_i)$, and the *state* is memoryless in the sense that the sequence $(S_1, S_2, \ldots)$ is independent and identically distributed (i.i.d.) with $S_i \sim p_S(s_i)$.

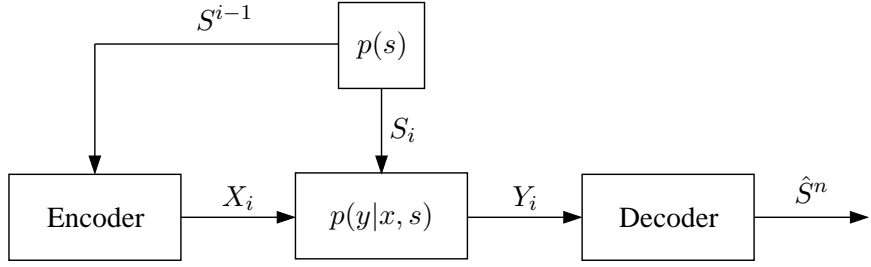

Fig. 1. Strictly causal state communication.

An $(|\mathcal{S}|^n, n)$ code for strictly causal state communication over the DMC with DM state consists of

- an encoder that assigns a symbol $x_i(s^{i-1}) \in \mathcal{X}$ to each past state sequence $s^{i-1} \in \mathcal{S}^{i-1}$ for $i \in [1:n]$, and
- a decoder that assigns an estimate $\hat{s}^n \in \hat{\mathcal{S}}^n$ to each received sequence $y^n \in \mathcal{Y}^n$.

The fidelity of the state estimate is measured by the *expected distortion*

$$\mathsf{E}(d(S^n, \hat{S}^n)) = \frac{1}{n} \sum_{i=1}^{n} \mathsf{E}(d(S_i, \hat{S}_i)),$$

where $d : \mathcal{S} \times \hat{\mathcal{S}} \to [0, \infty)$ is a distortion measure between a state symbol $s \in \mathcal{S}$ and a reconstruction symbol $\hat{s} \in \hat{\mathcal{S}}$. Without loss of generality, we assume that for every symbol $s \in \mathcal{S}$ there exists a reconstruction symbol $\hat{s} \in \hat{\mathcal{S}}$ such that $d(s, \hat{s}) = 0$.

A distortion $D$ is said to be *achievable* if there exists a sequence of $(|\mathcal{S}|^n, n)$ codes such that

$$\limsup_{n \to \infty} \mathsf{E}(d(S^n, \hat{S}^n)) \leq D.$$

We next characterize the minimum distortion $D^*$, which is the infimum of all achievable distortions $D$.

*Theorem 1:* The minimum distortion for strictly causal state communication is

$$D^* = \min \mathsf{E}(d(S, \hat{S})),$$

where the minimum is over all conditional pmfs $p(x)p(u|x,s)$ and functions $\hat{s}(u, x, y)$ such that

$$I(U, X; Y) \geq I(U, X; S).$$

To illustrate this result, we consider the following.

*Example 1 (Quadratic Gaussian state communication):* Consider the Gaussian channel with additive Gaussian state [5]

$$Y = X + S + Z,$$

where the state $S \sim \mathrm{N}(0, Q)$ and the noise $Z \sim \mathrm{N}(0, N)$ are independent. Assume an expected average transmission power constraint

$$\sum_{i=1}^{n} \mathsf{E}(x_i^2(S^{i-1})) \leq nP,$$

where the expectation is over the random state sequence $S^n$. We assume the *squared error (quadratic) distortion measure* $d(s, \hat{s}) = (s - \hat{s})^2$.

We compare different transmission strategies for estimating the state at the decoder. In the classical communication paradigm, the encoder would its ignore knowledge of the channel state (since the strictly causal state information at the encoder does not increase the channel capacity) and transmit an agreed-upon training sequence to the decoder. The minimum distortion is achieved by estimating the state $S_i$ via minimum mean squared error (MMSE) estimation from the noisy observation $Y_i = X_i + S_i + Z_i$ and

$$D = \mathsf{E}((\hat{S}_i - S_i)^2) = \mathsf{E}(S_i^2) - \frac{\mathsf{E}(S_i Y_i)^2}{\mathsf{E}(Y_i^2)} = \frac{QN}{Q+N}.$$

Note that the result is independent of the particular sequence $X_i$, i.e., one could "send" $X_i = 0, i \in [1:n]$. This distortion is optimal when the encoder is oblivious of the state sequence as shown in [31].

Alternatively, a block Markov coding scheme can be performed, in which the encoder communicates a description of the state sequence in the previous block using a capacity-achieving code. This strategy is similar to a source–channel separation scheme, whereby the state sequence is treated as a source and the compressed version of the source is sent across the noisy channel at a rate lower than the capacity. Since the distortion–rate function of the state is $D(R) = Q2^{-2R}$ (see, for example, [8]) and the capacity of the channel (with strictly causal state information at the encoder) is $C = \mathsf{C}(P/(Q+N))$, the distortion achieved by this coding scheme is $D = D(C) = Q(Q+N)/(P+Q+N)$. It is straightforward to see that for the same values of $P, Q$ and $N$, ignoring the state knowledge at the encoder can offer a lower distortion than using this (suboptimal) block Markov encoding scheme.

The minimum distortion however can be achieved again by performing another block Markov coding scheme, but this time the encoder communicates a description of the state sequence in the previous block by incorporating side information $(X, Y)$ about the state $S$ of previous block at the decoder. This strategy is equivalent to setting $X = \alpha U \sim \mathrm{N}(0, P)$, $U = S + \tilde{S}$, where $\tilde{S} \sim \mathrm{N}(0, Q/P)$ is independent of $(S, X)$, and $\hat{S} = \mathsf{E}(S|U, X, Y) = \mathsf{E}(S|S+\tilde{S}, S+Z)$ in Theorem 1. This strategy yields the minimum distortion given by $D^* = QN/(P+Q+N)$. (The proof of optimality is given in Section III.) This strategy, in effect, replaces $D(R) = Q2^{-2R}$ of the last scheme with the Wyner–Ziv distortion–rate function $D_{\mathrm{WZ}}(R) = (QN/(Q+N))2^{-2R}$ (see [30]) and the minimum distortion $D^*$ can be evaluated by computing $D_{\mathrm{WZ}}(C)$.

In the following two subsections, we prove Theorem 1.

## A. Proof of Achievability

We use $b$ transmission blocks, each block consisting of $n$ symbols. In block $j$, a description of the state sequence $S^n(j-1)$ in block $j-1$ is sent.

**Codebook generation.** Fix a conditional pmf $p(x)p(u|x,s)$ and function $\hat{s}(u,x,y)$ such that $I(U,X;Y) > I(U,X;S)$, and let $p(u|x) = \sum_s p(s)p(u|x,s)$. For each $j \in [1:b]$, randomly and independently generate $2^{nR_S}$ sequences $x^n(l_{j-1})$, $l_{j-1} \in [1:2^{nR_S}]$, each according to $\prod_{i=1}^n p_X(x_i)$. For each $l_{j-1} \in [1:2^{nR_S}]$, randomly and conditionally independently generate $2^{n\tilde{R}_S}$ sequences $u^n(k_j|l_{j-1})$, $k_j \in [1:2^{n\tilde{R}_S}]$, each according to $\prod_{i=1}^n p_{U|X}(u_i|x_i(l_{j-1}))$. Partition the set of indices $k_j \in [1:2^{n\tilde{R}_S}]$ into equal-size bins $\mathcal{B}(l_j) = [(l_j-1)2^{n(\tilde{R}_S-R_S)}+1 : l_j 2^{n(\tilde{R}_S-R_S)}]$, $l_j \in [1:2^{nR_S}]$. This defines the codebook

$$\mathcal{C}_j = \big\{(x^n(l_{j-1}), u^n(k_j|l_{j-1})\colon l_{j-1} \in [1:2^{nR_S}],\ k_j \in [1:2^{n\tilde{R}_S}]\big\}, \quad j \in [1:b].$$

The codebook is revealed to both the encoder and the decoder.

**Encoding.** By convention, let $l_0 = 1$. At the end of block $j$, the encoder finds an index $k_j$ such that

$$(s^n(j), u^n(k_j|l_{j-1}), x^n(l_{j-1})) \in \mathcal{T}_{\epsilon'}^{(n)}.$$

If there is more than one such index, it selects one of them uniformly at random. If there is no such index, it selects an index from $[1:2^{n\tilde{R}_S}]$ uniformly at random. In block $j+1$ the encoder transmits $x^n(l_j)$, where $l_j$ is the bin index of $k_j$.

**Decoding.** Let $\epsilon > \epsilon'$. At the end of block $j+1$, the decoder finds the unique index $\hat{l}_j$ such that $(x^n(\hat{l}_j), y^n(j+1)) \in \mathcal{T}_\epsilon^{(n)}$. (If there is more than one such index, it selects one of them uniformly at random. If there is no such index, it selects an index from $[1:2^{nR_S}]$ uniformly at random.) It then finds the unique index $\hat{k}_j \in \mathcal{B}(\hat{l}_j)$ such that $(u^n(\hat{k}_j|\hat{l}_{j-1}), x^n(\hat{l}_{j-1}), y^n(j)) \in \mathcal{T}_\epsilon^{(n)}$. Finally it computes the reconstruction sequence as $\hat{s}_i(j) = \hat{s}(u_i(\hat{k}_j|\hat{l}_{j-1}), x_i(\hat{l}_{j-1}), y_i(j))$ for $i \in [1:n]$.

**Analysis of expected distortion.** Let $L_{j-1}, K_j, L_j$ be the indices chosen in block $j$. We bound the distortion averaged over the random choice of the codebooks $\mathcal{C}_j$, $j \in [1:b]$. Define the "error" event

$$\mathcal{E}(j) = \big\{(S^n, U^n(\hat{K}_j|\hat{L}_{j-1}), X^n(\hat{L}_{j-1}), Y^n(j)) \notin \mathcal{T}_\epsilon^{(n)}\big\}$$

and consider the events

$$\mathcal{E}_1(j) = \big\{(S^n, U^n(K_j|L_{j-1}), X^n(L_{j-1}), Y^n(j)) \notin \mathcal{T}_\epsilon^{(n)}\big\},$$

$$\mathcal{E}_2(j-1) = \{\hat{L}_{j-1} \neq L_{j-1}\},$$

$$\mathcal{E}_2(j) = \{\hat{L}_j \neq L_j\},$$

$$\mathcal{E}_3(j) = \{\hat{K}_j \neq K_j\}.$$

Then by the union of events bound,

$$\mathsf{P}\{\mathcal{E}(j)\} \leq \mathsf{P}\{\mathcal{E}_1(j)\} + \mathsf{P}\{\mathcal{E}_2(j-1)\} + \mathsf{P}\{\mathcal{E}_2(j)\} + \mathsf{P}\{\mathcal{E}_2^c(j-1) \cap \mathcal{E}_2^c(j) \cap \mathcal{E}_3(j)\}.$$

We bound each term. For the first term, let

$$\tilde{\mathcal{E}}_1(j) = \big\{(S^n, U^n(K_j|L_{j-1}), X^n(L_{j-1})) \notin \mathcal{T}_{\epsilon'}^{(n)}\big\}$$

and note that

$$\mathsf{P}\{\mathcal{E}_1(j)\} \leq \mathsf{P}\{\tilde{\mathcal{E}}_1(j)\} + \mathsf{P}\{\tilde{\mathcal{E}}_1^c(j) \cap \mathcal{E}_1(j)\}.$$

By the independence of the codebooks (in particular, the independence of $L_{j-1}$ and $\mathcal{C}_j$) and the covering lemma [11, Sec. 3.7], $\mathsf{P}\{\tilde{\mathcal{E}}_1(j)\}$ tends to zero as $n \to \infty$ if $\tilde{R}_S > I(U;S|X) + \delta(\epsilon')$. Since $\epsilon > \epsilon'$ and $Y^n(j)|\{U^n(K_j|L_{j-1} = u^n, X^n(L_{j-1}) = x^n, S^n(j) = s^n\} \sim \prod_{i=1}^n p_{Y|X,S}(y_i|x_i, s_i)$, by the conditional typicality lemma [11, Sec. 2.5], $\mathsf{P}\{\tilde{\mathcal{E}}_1^c(j) \cap \mathcal{E}_1(j)\}$ tends to zero as $n \to \infty$.

Next, by the same independence of the codebooks and the packing lemma [11, Sec. 3.2], $\mathsf{P}\{\mathcal{E}_2(j-1)\}$ and $\mathsf{P}\{\mathcal{E}_2(j)\}$ tend to zero as $n \to \infty$ if $R_S < I(X;Y) - \delta(\epsilon)$. Finally, following the same steps as in the analysis of the Wyner–Ziv coding scheme [11, Sec. 11.3] (in particular, the analysis of $\mathcal{E}_3$), it can be readily shown that $\mathsf{P}\{\mathcal{E}_2^c(j-1) \cap \mathcal{E}_2^c(j) \cap \mathcal{E}_3(j)\}$ tends to zero as $n \to \infty$ if $\tilde{R}_S - R_S < I(U;Y|X) - \delta(\epsilon)$. Combining the bounds and eliminating $\tilde{R}_S$ and $R_S$, we have shown that $\mathsf{P}\{\mathcal{E}(j)\}$ tends to zero as $n \to \infty$ if $I(U,X;Y) > I(U;S|X) + \delta(\epsilon') + 2\delta(\epsilon) = I(U,X;S) + \delta'(\epsilon)$, which is satisfied by our choice of $p(x)p(u|x,s)$ for $\epsilon$ sufficiently small.

When there is no "error" $(S^n, U^n(\hat{K}_j|\hat{L}_{j-1}), X^n(\hat{L}_{j-1}), Y^n(j)) \in \mathcal{T}_\epsilon^{(n)}$. Thus, by the law of total expectation and the typical average lemma [11, Sec. 2.4], the asymptotic distortion averaged over the random codebook, encoding, and decoding is upper bounded as

$$\limsup_{n \to \infty} \mathsf{E}(d(S^n(j), \hat{S}^n(j))) \leq \limsup_{n \to \infty} \big(d_{\max} \mathsf{P}\{\mathcal{E}(j)\} + (1+\epsilon)\mathsf{E}(d(S,\hat{S}))\mathsf{P}\{\mathcal{E}^c(j)\}\big)$$

$$\leq (1+\epsilon)\mathsf{E}(d(S,\hat{S})),$$

where $d_{\max} = \max_{(s,\hat{s}) \in \mathcal{S} \times \hat{\mathcal{S}}} d(s,\hat{s}) < \infty$. By taking $\epsilon \to 0$ and $b \to \infty$, any distortion larger than $\mathsf{E}(d(S,\hat{S}))$ is achievable for a fixed conditional pmf $p(x)p(u|x,s)$ and function $\hat{s}(u,x,y)$ satisfying $I(U,X;Y) > I(U,X;S)$. Finally, by the continuity of mutual information terms in $p(x)p(u|x,s)$, the same conclusion holds when we relax the strict inequality to $I(U,X;Y) \geq I(U,X;S)$. This completes the achievability proof of Theorem 1.

## B. Proof of the Converse

In this section, we prove that for every code, the achieved distortion is lower bounded as $D \geq D^*$. Given an $(|\mathcal{S}|^n, n)$ code, we identify the auxiliary random variables $U_i = (S^{i-1}, Y_{i+1}^n)$, $i \in [1:n]$. Note that, as desired, $U_i \to (X_i, S_i) \to Y_i$ form a Markov chain for $i \in [1:n]$. Consider

$$\sum_{i=1}^n I(U_i, X_i; S_i) = \sum_{i=1}^n I(S^{i-1}, Y_{i+1}^n, X_i; S_i)$$

$$\stackrel{(a)}{=} \sum_{i=1}^n I(S^{i-1}, Y_{i+1}^n; S_i)$$

$$= \sum_{i=1}^n \bigl(I(S^{i-1}; S_i) + I(Y_{i+1}^n; S_i | S^{i-1})\bigr)$$

$$\stackrel{(b)}{=} \sum_{i=1}^n I(Y_{i+1}^n; S_i | S^{i-1})$$

$$\stackrel{(c)}{=} \sum_{i=1}^n I(S^{i-1}; Y_i | Y_{i+1}^n)$$

$$\leq \sum_{i=1}^n I(S^{i-1}, Y_{i+1}^n, X_i; Y_i)$$

$$= \sum_{i=1}^n I(U_i, X_i; Y_i), \qquad (1)$$

where $(a)$ follows since $X_i$ is a function of $S^{i-1}$, $(b)$ follows since $S^n$ is i.i.d., and $(c)$ follows by the Csiszár sum identity [9], [13], [11, Sec. 2.3].

Let $Q$ be a time-sharing random variable, uniformly distributed over $[1:n]$ and independent of $(X^n, S^n, Y^n)$, and let $U = (Q, U_Q)$, $X = X_Q$, $S = S_Q$, and $Y = Y_Q$. It can be easily verified that $X$ is independent of $S$ and $U \to (X, S) \to Y$ form a Markov chain. Furthermore

$$I(U, X; S) \stackrel{(a)}{=} I(U_Q, X_Q; S_Q | Q)$$

$$= \frac{1}{n} \sum_{i=1}^{n} I(U_i, X_i; S_i)$$

$$\stackrel{(b)}{\leq} \frac{1}{n} \sum_{i=1}^{n} I(U_i, X_i; Y_i)$$

$$= I(U_Q, X_Q; Y_Q | Q)$$

$$\leq I(U, X; Y),$$

where $(a)$ follows since $Q$ is independent of $S_Q$ and $(b)$ follows from the definition of the code.

To lower bound the expected distortion of the given code, we rely on the following result.

*Lemma 1:* Suppose $Z \to V \to W$ form a Markov chain and $d(z, \hat{z})$ is a distortion measure. Then for every reconstruction function $\hat{z}(v, w)$, there exists a reconstruction function $\hat{z}^*(v)$ such that

$$\mathsf{E}\big[d(Z, \hat{z}^*(V))\big] \leq \mathsf{E}\big[d(Z, \hat{z}(V, W))\big].$$

This extremely useful lemma traces back to Blackwell's notion of channel ordering [1], [25] and can be interpreted as a "data processing inequality" for estimation. In the context of network information theory, it has been utilized by Kaspi [15] (see also [11, Section 20.3.3]) and appeared in the above simple form in [3]. For completeness, the proof of this lemma is provided in Appendix A.

Now consider

$$\mathsf{E}\big[d(S^n, \hat{S}^n)\big] = \frac{1}{n} \sum_{i=1}^{n} E\big[d(S_i, \hat{s}_i(Y^n))\big]$$

$$\stackrel{(a)}{\geq} \frac{1}{n} \sum_{i=1}^{n} \min_{\hat{s}^*(i, u_i, x_i, y_i)} \mathsf{E}\big[d(S_i, \hat{s}^*(i, U_i, X_i, Y_i))\big]$$

$$= \min_{\hat{s}^*(u, x, y)} E\big[d(S, \hat{s}^*(U, X, Y))\big],$$

where $(a)$ follows from Lemma 1 by identifying $S_i$ as $Z$, $(U_i, X_i, Y_i) = (S^{i-1}, X_i, Y_i^n)$ as $V$, and $Y^{i-1}$ as $W$, and noting that $S_i \to (S^{i-1}, X_i, Y_i^n) \to Y^{i-1}$ form a Markov chain. This completes the proof of Theorem 1.

*C. Lossless Communication*

Suppose that the state sequence needs to be communicated *losslessly*, i.e., $\lim_{n \to \infty} \mathsf{P}\{\hat{S}^n \neq S^n\} = 0$. We can establish the following congruence of Theorem 1.

*Corollary 1:* If $H(S) < \Delta^* = \max_{p(x)} I(X,S;Y)$, then the state sequence can be communicated losslessly. Conversely, if the state sequence can be communicated losslessly, then $H(S) \le \Delta^*$.

To prove this, consider the special case of $\hat{S} = S$ and Hamming distortion measure $d(s,\hat{s})$ (i.e., $d(s,\hat{s}) = 0$ if $s = \hat{s}$ and 1 if $s \ne \hat{s}$). By setting $U = S$ in the achievability proof of Theorem 1 in Subsection II-A and noting that no "error" implies that $S^n = \hat{S}^n$, we can conclude that the state sequence can be communicated losslessly if $\Delta^* > H(S)$ for some $p(x)$. The converse follows immediately since the lossless condition that the block error probability $\mathsf{P}\{\hat{S}^n \ne S^n\}$ tends to zero as $n \to \infty$ implies the zero Hamming distortion condition that the average symbol error probability $(1/n)\sum_{i=1}^n \mathsf{P}\{\hat{S}_i \ne S_i\}$ tends to zero as $n \to \infty$. Combining this observation with the converse proof of Theorem 1 in Subsection II-B, we can conclude that $H(S)$ must be less than or equal to $\Delta^*$.

*Remark 1:* If we define $\Delta^* = \max_{p(x)} I(X,S;Y)$, then $\min\{H(S), \Delta^*\}$ characterizes the *state uncertainty reduction rate,* which captures the performance of the optimal list decoder for the state sequence (see [17] for the exact definition). The proof of this result again follows from Theorem 1 by letting $\hat{S}$ be the set of pmfs on $S$ and $d(s,\hat{s}) = \log(1/\hat{s}(s))$ be the logarithmic distortion measure and adapting the technique by Courtade and Weissman [6].

## III. CAPACITY–DISTORTION TRADEOFF

Now suppose that in addition to the state sequence $S^n$, the encoder wishes to communicate a message $M$ independent of $S^n$. What is the optimal tradeoff between the rate $R$ of the message and the distortion $D$ of state estimation?

A $(2^{nR}, n)$ code for strictly causal state communication consists of

- a message set $[1:2^{nR}]$,
- an encoder that assigns a symbol $x_i(m, s^{i-1}) \in \mathcal{X}$ to each message $m \in [1:2^{nR}]$ and past state sequence $s^{i-1} \in \mathcal{S}^{i-1}$ for $i \in [1:n]$, and
- a decoder that assigns a message estimate $\hat{m} \in [1:2^{nR}]$ (or an error message e) and a state sequence estimate $\hat{s}^n \in \hat{\mathcal{S}}^n$ to each received sequence $y^n \in \mathcal{Y}^n$.

We assume that $M$ is uniformly distributed over the message set. The average probability of error is defined as $P_e^{(n)} = \mathsf{P}\{\hat{M} \ne M\}$. As before, the channel state estimation error is defined as $\mathsf{E}(d(S^n, \hat{S}^n))$. A rate–distortion pair is said to be achievable if there exists a sequence of $(2^{nR}, n)$ codes such that $\lim_{n \to \infty} P_e^{(n)} = 0$ and $\limsup_{n \to \infty} \mathsf{E}(d(S^n, \hat{S}^n)) \le D$. The capacity–distortion function $C_{\mathrm{SC}}(D)$ is the supremum of the rates $R$ such that $(R, D)$ is achievable.

We characterize this optimal tradeoff between information transmission rate (capacity $C$) and state estimation (distortion $D$) as follows.

*Theorem 2:* The capacity–distortion function for strictly causal state communication is

$$C_{\text{SC}}(D) = \max \bigl( I(U, X; Y) - I(U, X; S) \bigr), \tag{2}$$

where the maximum is over all conditional pmfs $p(x)p(u|x,s)$ with $|\mathcal{U}| \leq |\mathcal{S}|+2$ and functions $\hat{s}(u,x,y)$ such that $\mathsf{E}(d(S,\hat{S})) \leq D$.

The proof of Theorem 2 is similar to the zero-rate case in Theorem 1 and thus we delegate it to Appendix B. Note that the inverse of the capacity–distortion function, namely, the distortion–capacity function for strictly causal state communication is

$$D_{\text{SC}}(C) = \min \mathsf{E}(d(S,\hat{S})), \tag{3}$$

where the minimum is over all conditional pmfs $p(x)p(u|x,s)$ and functions $\hat{s}(u,x,y)$ such that $I(U,X;Y) - I(U,X;S) \geq C$. By setting $C = 0$ in (3), we recover Theorem 1. (More interestingly, we can recover Theorem 2 from Theorem 1 by considering a supersource $S' = (S, W)$, where the message source $W$ is independent of $S$, and two distortion measures—the Hamming distortion measure $d(w, \hat{w})$ and a generic distortion measure $d(s, \hat{s})$.) At the other extreme, by setting $D = \infty$ in (2), we recover the capacity expression

$$C = \max_{p(x)} I(X;Y) \tag{4}$$

of a DMC with DM state when the state information is available strictly causally at the encoder. (Unlike the general tradeoff in Theorem 2, strictly causal state information is useless when communicating the message alone.) Finally, by setting $U = \emptyset$ in Theorem 2, we recover the result in [31] on the capacity–distortion function when the state information is not available at the encoder.

*Remark 2:* Theorem 2 (as well as Theorem 1) holds for any finite delay, that is, whenever the encoder is defined as $x_i(m, s^{i-d})$ for some $d \in [1:\infty)$. More generally, it continues to hold as long as the delay is sublinear in the block length $n$.

*Remark 3:* The characterization of the capacity–distortion function in Theorem 2, albeit very compact, does not bring out the intrinsic tension between state estimation and independent information transmission. It can be alternatively written as

$$C_{\text{SC}}(D) = \max_{p(x), D_x : \mathsf{E}_X(D_X) \leq D} \bigl( I(X;Y) - \mathsf{E}_X[R_{\text{WZ}}^{(X)}(D_X)] \bigr), \tag{5}$$

October 29, 2018                                                                                                                                    DRAFT

where
$$R_{\text{WZ}}^{(x)}(D) = \min_{p(u|x,s), \hat{s}(u,x,y): \mathsf{E}[d(S,\hat{S}(U,x,Y))] \leq D} I(U; S | x, Y), \quad x \in \mathcal{X},$$

is the Wyner–Ziv rate–distortion function with side information $(x, Y)$. The rate $R_{\text{WZ}}^{(x)}(D_x)$ can be viewed as the price the encoder pays to estimate the channel state at the decoder under distortion $D_x$ by signaling with $x$. In particular, if $R_{\text{WZ}}^{(x)}(D)$ is independent of $x$ for a fixed $D$ (i.e., $R_{\text{WZ}}^{(x)}(D) = R_{\text{WZ}}(D)$), then by the convexity of the Wyner–Ziv rate–distortion function, the alternative characterization of $C_{\text{SC}}(D)$ in (5) simplifies to

$$C_{\text{SC}}(D) = C_{\text{SC}}(\infty) - R_{\text{WZ}}(D), \tag{6}$$

where

$$R_{\text{WZ}}(D) = R_{\text{WZ}}^{(x)}(D), \quad x \in \mathcal{X}.$$

Thus, in this case the capacity is achieved by splitting the unconstrained capacity $C_{\text{SC}}(\infty)$ into information transmission and lossy source coding of the past state sequence with side information $(X, Y)$. This simple characterization will be very useful in evaluating the capacity–distortion function in several examples.

*Remark 4:* Along the same lines of [17], the optimal tradeoff between the state uncertainty reduction rate $\Delta$ and independent information transmission rate $R$ can be characterized as the set of $(R, \Delta)$ pairs such that

$$R \leq I(X; Y)$$

$$\Delta \leq H(S)$$

$$R + \Delta \leq I(X, Y; S)$$

for some $p(x)$. This result includes both the state uncertainty reduction rate in Remark 1 and the channel capacity in (4) as special cases.

In the following subsections, we illustrate Theorem 2 via simple examples.

### A. Injective Deterministic Channels

Suppose that the channel output

$$Y = y(X, S)$$

is a function of $X$ and $S$ such that given every $x \in \mathcal{X}$, the function $y(x, s)$ is injective (one-to-one) in $s$. This condition implies that $H(Y|X) = H(S)$ for every $p(x)$. For this class of injective deterministic channels, the characterization of the capacity–distortion function in Theorem 2 can be greatly simplified.

*Proposition 1:* The capacity–distortion function of the injective deterministic channel is

$$C_{\text{SC}}(D) = C_{\text{SC}}(0) = \max_{p(x)} I(X;Y) = \max_{p(x)} \big(H(Y) - H(S)\big). \tag{7}$$

In other words, we can achieve the unconstrained channel capacity as well as perfect state estimation. This is no surprise since the injective condition implies that given the channel input $X$ and output $Y$, the state $S$ can be recovered losslessly. Note that this result is independent of the distortion measure $d(s, \hat{s})$ as long as our critical assumption—for every $s$, there exists an $\hat{s}$ with $d(s, \hat{s}) = 0$—is satisfied.

To prove achievability in Proposition 1, substitute $U = Y$ in Theorem 2. For the converse, consider

$$\begin{aligned}
I(U, X; Y) - I(U, X; S) &= I(X;Y) - \big(I(U;S|X) - I(U;Y|X)\big) \\
&= I(X;Y) - \big(H(U|Y,X) - H(U|X,S)\big) \\
&\stackrel{(a)}{=} I(X;Y) - (H(U|Y,X) - H(U|Y,X,S)) \\
&= I(X;Y) - I(U;S|Y,X) \\
&\stackrel{(b)}{=} I(X;Y),
\end{aligned}$$

where $(a)$ follows since $Y = y(X, S)$ and $(b)$ follows from the injective condition.

*Example 2 (Gaussian channel with additive Gaussian state and no noise):* Consider the channel

$$Y = X + S,$$

where the state $S \sim \text{N}(0, Q)$. Assume the squared error distortion measure and an expected average power constraint $P$ on $X$. The capacity–distortion function of this channel is

$$C_{\text{SC}}(D) = \mathsf{C}(P/Q) \quad \text{for all } D,$$

which is the capacity without state estimation.

*Example 3 (Binary symmetric channel with additive Bernoulli state and no noise):* Consider the channel

$$Y = X \oplus S,$$

where $X$ and $Y$ are binary and the state $S \sim \text{Bern}(q)$. Assume the Hamming distortion measure. The capacity–distortion function of this channel is

$$C_{\text{SC}}(D) = 1 - H(q) \quad \text{for all } D.$$

In the following subsections, we extend the above two examples to the more general cases where there is additive noise.

## B. Gaussian Channel with Additive Gaussian State

We revisit the Gaussian channel with additive Gaussian noise (see Example 1)

$$Y = X + S + Z,$$

where $S \sim \text{N}(0, Q)$ and $Z \sim \text{N}(0, N)$. As before, we assume an average expected power constraint $P$ and the squared error distortion measure $d(x, \hat{x}) = (x - \hat{x})^2$.

We note the following extreme cases of the capacity–distortion function:

- If $N = 0$, then $C_{\text{SC}}(D) = C_{\text{SC}}(\infty) = \infty$.
- If $D \leq D^* = QN/(P+Q+N)$ (the optimal distortion mentioned in Example 1), then $C_{\text{SC}}(D) = 0$.
- If $D \geq QN/(Q+N)$ (the minimum distortion achievable when the encoder has no knowledge of the state), then $C(D) = C(\infty) = \mathsf{C}(P/(Q+N))$, which is achieved by first decoding the codeword $X^n$ in a "noncoherent" fashion, then utilizing $X^n$ along with the channel output $Y^n$ to estimate $S^n$ (see [31]).

More generally, we have the following.

*Proposition 2:* The capacity–distortion function of the Gaussian channel with additive Gaussian state when the state information is strictly causally available at the encoder is

$$C_{\text{SC}}(D) = \begin{cases} 0, & 0 \leq D < \frac{QN}{P+Q+N}, \\ \mathsf{C}\left(\frac{(P+Q+N)D - QN}{QN}\right), & \frac{QN}{P+Q+N} \leq D < \frac{QN}{Q+N}, \\ \mathsf{C}\left(\frac{P}{Q+N}\right), & D \geq \frac{QN}{Q+N}. \end{cases}$$

Proposition 2 can be proved by evaluating the characterization in Theorem 2 with the optimal choice of the auxiliary random variable $U$ and the estimation function $\hat{s}(u, x, y)$. However, the alternative characterization in Remark 3 provides a more direct proof. Since the Wyner–Ziv rate–distortion function [30] for the Gaussian source $S$ with side information $Y = x + S + Z$ is independent of $x$, it follows immediately from (6) that $C_{\text{SC}}(D) = C_{\text{SC}}(\infty) - R_{\text{WZ}}(D)$, which is equivalent to the expression given in Proposition 2.

## C. Binary Symmetric Channel with Additive Bernoulli State

Consider the binary symmetric channel

$$Y = X \oplus S \oplus Z,$$

where the state $S \sim \text{Bern}(q)$, $q \in [0, 1/2]$, and the noise $Z \sim \text{Bern}(p)$, $p \in [0, 1/2]$, are independent of each other. Assume the Hamming distortion measure $d(x, \hat{x}) = x \oplus \hat{x}$.

We note the following extreme cases of the capacity–distortion function:

- If $p = 0$, then $D^* = 0$ and $C_{\text{SC}}(D) \equiv 1 - H(q)$.
- If $q = 0$, then $D^* = 0$ and $C_{\text{SC}}(D) \equiv 1 - H(p)$.
- If $p = 1/2$, then $D^* = q$ and $C_{\text{SC}}(D) \equiv 0$.
- If $q = 1/2$, then $D^* = p$ and $C_{\text{SC}}(D) \equiv 0$.
- If $D \geq q$, then $C_{\text{SC}}(D) = C_{\text{SC}}(\infty) = 1 - H(p * q) = 1 - H(p(1-q) + q(1-p))$.

More generally, we have the following.

*Proposition 3:* The capacity–distortion function of the binary symmetric channel with additive Bernoulli state when the state information is strictly causally available at the encoder is

$$C_{\text{SC}}(D) = \max_{\alpha,\beta \in [0,1]:\, \alpha\beta+(1-\alpha)q \leq D} \left[1 - H(p) - \alpha\big(H(\beta * q) - H(\beta)\big)\right]$$

$$= 1 - H(p * q) - R_{\text{WZ}}(D),$$

where

$$R_{\text{WZ}}(D) = \min_{\alpha,\beta \in [0,1]:\, \alpha\beta+(1-\alpha)q \leq D} \left[H(p) - H(p * q) + \alpha\big(H(\beta * q) - H(\beta)\big)\right] \tag{8}$$

is the Wyner–Ziv rate-distortion function for the Bernoulli source and Hamming distortion measure.

As in the Gaussian case, the proof of the proposition follows immediately from the alternative characterization of the capacity–distortion function in Remark 3. Here the Wyner–Ziv rate–distortion function follows again from [30].

## IV. Causal State Communication

So far in our discussion, we have assumed that the encoder has strictly causal knowledge of the state sequence. What will happen if the encoder has *causal* knowledge of the state sequence, that is, at time $i \in [1:n]$ the previous and current state sequence $s^i$ is available at the encoder? Now a $(2^{nR}, n)$ code, probability of error, achievability, and capacity–distortion function are defined as in the strictly causal case in Section III, except that the encoder is of the form $x_i(m, s^i), i \in [1:n]$.

It turns out that the optimal tradeoff between capacity and distortion can be achieved by a simple modification to the block Markov coding scheme for the strictly causal case.

*Theorem 3:* The capacity–distortion function for causal state communication is

$$C_{\text{C}}(D) = \max\big(I(U,V;Y) - I(U,V;S)\big), \tag{9}$$

where the maximum is over all conditional pmfs $p(v)p(u|v,s)$ with $|\mathcal{V}| \leq \min\{(|\mathcal{X}|-1)|\mathcal{S}|+1, |\mathcal{Y}|\}+1$ and $|\mathcal{U}| \leq |\mathcal{S}|+2$ and functions $x(v,s)$ and $\hat{s}(u,v,y)$ such that $\mathsf{E}(d(S,\hat{S})) \leq D$.

October 29, 2018 DRAFT

At one extreme point, if $D = \infty$, then the theorem recovers the unconstrained channel capacity

$$C_{\text{C}}(\infty) = \max_{p(v)p(u|v,s),\, x(v,s)} \big(I(U,V;Y) - I(U,V;S)\big) = \max_{p(v),\, x(v,s)} I(V;Y)$$

established by Shannon [26]. At the other extreme point, the optimal distortion for causal state communication is

$$D^* = \min \mathsf{E}(d(S, \hat{S})),$$

where the minimum is over all conditional pmfs $p(v)p(u|v,s)$ and functions $x(v,s)$ and $\hat{s}(u,v,y)$ such that

$$I(U,V;Y) \geq I(U,V;S).$$

Moreover, the condition for zero Hamming distortion can be shown to be

$$\max_{p(x|s)} I(X,S;Y) \geq H(S),$$

which was proved in [17]. Note that by setting $V = X$ in the theorem, we recover the capacity–distortion function $C_{\text{SC}}(D)$ for strictly causal communication in Theorem 2.

To prove achievability for Theorem 3, we use the Shannon strategy [26] (see also [11, Sec. 7.5]) and perform encoding over the set of all functions $\{x_v(s)\colon \mathcal{S} \mapsto \mathcal{X}\}$ indexed by $v$ as the input alphabet. This induces a DMC with DM state $p(y|v,s)p(s) = p(y|x(v,s),s)p(s)$ with the state information strictly causally available at the encoder and we can immediately apply Theorem 2 to prove achievability of $C_{\text{C}}(D)$. For the converse, we identify the auxiliary random variables $V_i = (M, S^{i-1})$ and $U_i = Y_{i+1}^n$, $i \in [1:n]$. Note that $(U_i, V_i) \to (X_i, S_i) \to Y_i$ form a Markov chain, $V_i$ is independent of $S_i$, and $X_i$ is a function of $(V_i, S_i)$ as desired. The rest of the proof utilizes Lemma 1 and the concavity of $C_{\text{C}}(D)$, and follows similar steps to that for the strictly causal case in Appendix B.

In the following subsections, we illustrate Theorem 3 through simple examples.

*A. Gaussian Channel with Additive Gaussian State*

We revisit the Gaussian channel (see Example 1 and Subsection III-B)

$$Y = X + S + Z.$$

While the complete characterization of $C_{\text{C}}(D)$ is not known even for the unconstrained case ($D = \infty$), the optimal distortion can be characterized as

$$D^* = \frac{QN}{\left(\sqrt{P} + \sqrt{Q}\right)^2 + N}.$$

Achievability follows by setting $U = V = \emptyset$, $X = \sqrt{P/Q}\,S$, and $\hat{S} = \mathsf{E}(S|Y)$. The converse follows from the fact that $D^*$ is also the optimal distortion when the state information is known *noncausally* at the encoder (see [29]). It is evident that knowing channel state causally helps the encoder to coherently choose the channel codeword $X$ to amplify the channel state $S$ unlike the strictly causal case where $X$ and $S$ are independent of each other.

### B. Binary Symmetric Channel with Additive Bernoulli State

We revisit the binary symmetric channel (see Subsection III-C)

$$Y = X \oplus S \oplus Z,$$

where $S \sim \text{Bern}(q)$ and $Z \sim \text{Bern}(p)$ are independent of each other.

We note the following extreme cases of the capacity–distortion function:

- If $p = 0$, then $D^* = 0$ and $C_{\text{C}}(D) \equiv 1 - H(q)$.
- If $q = 0$, then $D^* = 0$ and $C_{\text{C}}(D) \equiv 1 - H(p)$.
- If $p = 1/2$, then $D^* = q$ and $C_{\text{C}}(D) \equiv 0$.
- If $D \geq q$, then $C_{\text{C}}(D) = C_{\text{C}}(\infty) = 1 - H(p)$, which is achieved by canceling the state at the encoder ($X = V \oplus S$).

In general, the capacity–distortion function is given by the following proposition.

*Proposition 4:* The capacity–distortion function of the binary symmetric channel with additive Bernoulli state when the state information is causally available at the encoder is

$$C_{\text{C}}(D) = 1 - H(p) - H(q) + H(D), \quad D \leq q.$$

*Proof:* For the proof of achievability, observe that if we cancel the state at the encoder and split the unconstrained capacity into information transmission and lossy source coding of the past state sequence (without side information since $V$ and $Y$ are independent of $S$), then $C_{\text{C}}(\infty) - R(D) = (1 - H(p)) - (H(q) - H(D))$ is achievable. This corresponds to evaluating Theorem 2 with $X = V \oplus S$, $U = V \oplus S \oplus \tilde{S}$, and $\hat{S} = U \oplus V = S \oplus \tilde{S}$, where $V \sim \text{Bern}(1/2)$ and $\tilde{S} \sim \text{Bern}(D)$ are independent of $S$. (Note the similarity to rate splitting for the strictly causal case discussed in Remark 3.)

For the proof of the converse, consider

$$I(U,V;Y) - I(U,V;S) = I(U,V,S;Y) - I(U,V,Y;S)$$

$$= H(Y) - H(Y|U,V,S) - H(S) + H(S|U,V,Y)$$

$$\stackrel{(a)}{=} H(Y) - H(Y|X,S) - H(S) + H(S|U,V,Y)$$

$$\stackrel{(b)}{=} H(Y) - H(Y|X,S) - H(S) + H(S \oplus \hat{S}|U,V,Y)$$

$$\leq 1 - H(p) - H(q) + H(S \oplus \hat{S})$$

$$= 1 - H(p) - H(q) + H(D),$$

where $(a)$ follows since $X$ is a function of $(V,S)$ and $(U,V) \to (X,S) \to Y$ form a Markov chain, and $(b)$ follows since $\hat{S}$ is a function of $(U,V,Y)$. This completes the proof of the proposition. ∎

## C. Five-Card Trick

We next consider the classical five-card trick. Two information theorists, Alice and Bob, perform a "magic" trick with a shuffled deck of $N$ cards, numbered from $0$ to $N-1$. Alice asks a member of the audience to select $K$ cards at random from the deck. The audience member passes the $K$ cards to Alice, who examines them and hands one back. Alice then arranges the remaining $K-1$ cards in some order and places them face down in a neat pile. Bob, who has not witnessed these proceedings, then enters the room, looks at the $K-1$ cards, and determines the missing $K$-th card, held by the audience member. There are two key questions:

- Given $K$, find the maximum number of cards $N$ for which this trick could be performed?
- How is this trick performed?

This trick (discussed in [7], [23]) can be formulated as state communication at zero Hamming distortion with causal state knowledge at the encoder.

*Proposition 5:* The maximum number of cards $N$ for which the trick could be performed is $K!+K-1$.

*Proof:* To show that the maximum cannot be larger than $K!+K-1$, that is, to prove the converse, we suppose that multiple rounds of the trick were to be performed. In the framework of causal state communication, the state $S$ corresponds to an unordered tuple of $K$ cards selected by the audience member, which is uniformly distributed over all possible choices of $K$ cards. The channel input $X$ (as well as the channel output $Y$) corresponds to the ordered tuple of $K-1$ cards placed and received, respectively, by Alice and Bob. Since Bob has to recover the missing card losslessly, the problem is equivalent to reproducing the state $S$ itself with zero Hamming distortion (by combining the remaining card with the received $K-1$ cards).

Now by Theorem 3, the necessary condition for zero Hamming distortion is given by

$$\max_{p(x|s)} H(X) - H(S) \geq 0,$$

or equivalently,

$$\max_{p(x|s)} \big(H(X|S) - H(S|X)\big) \geq 0. \tag{10}$$

Since $S$ is uniform and the maximum is attained by the (conditionally) uniform $X$, the condition in (10) simplifies to

$$\log(K!) \geq \log(N - (K-1)),$$

or equivalently,

$$N \leq K! + K - 1.$$

We now show that we only need one round of communication to achieve this upper bound on causal state communication. Without loss of generality, assume that the selected cards $(c_0, \cdots, c_{K-1})$ are ordered with $c_0 < c_1 < \cdots < c_{K-1}$. Alice selects card $c_i$ to hand back to the audience where $i = c_0 + c_1 + \cdots + c_{K-1} \pmod{K}$. Observe that

$$c_0 + c_1 + \cdots + c_{K-1} = Kr_1 + i, \tag{11}$$

for some integer $r_1$. The remaining $K-1$ cards $(c_{j_1}, \cdots, c_{j_{K-1}})$ ($c_{j_0} = c_i$ is the deleted card) are summed and decomposed, i.e.,

$$c_{j_1} + c_{j_2} + \cdots + c_{j_{K-1}} = Kr_2 + s, \tag{12}$$

for some integer $r_2$. Since all the $K$ cards sum to $i \pmod{K}$, the missing card $c_{j_0} = c_i$ must be congruent to $-s + i \pmod{K}$. Thus

$$c_{j_0} = c_i = K(r_1 - r_2) - s + i. \tag{13}$$

Therefore, if we renumber the $N - (K-1)$ cards from 0 to $K! - 1$ (by removing the $K-1$ retained cards), the hidden card's new number is congruent to $-s \pmod{K}$ as the hidden card's new number $c_i - i$ is equal to $K(r_1 - r_2) - s$. But there are exactly $(K-1)!$ possibilities remaining for the hidden card's number, which can be conveyed by a predetermined permutation of the $K-1$ retained cards. This completes the achievability proof. ∎

20## V. Concluding Remarks

The problem of joint information transmission and channel state estimation over a DMC with DM state was studied in [31] (no state information at the encoder) and [28], [29] (full state information at the encoder). In this paper, we bridged the temporal gap between these two results by studying the case in which the encoder has strictly causal or causal knowledge of the channel state information.

The resulting capacity–distortion function permits a systematic investigation of the tradeoff between information transmission and state estimation. We showed the use of block Markov coding coupled with channel state estimation by treating the decoded message and received channel output as side information at the decoder is optimal for communicating the state. Additional information transmission requires a simple rate-splitting strategy. We also showed that the capacity–distortion function when the encoder is oblivious of the state information (see [31]) can be recovered from our result.

Finally, we recall an important open problem of finding the capacity–distortion function $C_{\text{NC}}(D)$ for a general DMC with DM state with an arbitrary distortion measure, when the state sequence is *noncausally* at the encoder. The problem was studied in [28], which established a lower bound on $C_{\text{NC}}(D)$ as

$$C_{\text{NC}}(D) \geq \max\bigl(I(U;Y) - I(U;S)\bigr), \tag{14}$$

where the maximum is over all conditional pmfs $p(u|s)$ and functions $x(u,s)$ and $\hat{s}(u,y)$ such that $\mathsf{E}(d(S,\hat{S})) \leq D$. While it is believed that this lower bound is tight in general (see, for example, [29] for the case of Gaussian channels with additive Gaussian states with quadratic distortion measure), the proof of the converse seems beyond our current techniques of identifying auxiliary random variables and using estimation-theoretic inequalities such as Lemma 1.

## VI. Acknowledgments

The authors would like to thank Sung Hoon Lim for enlightening discussions.

## Appendix A
### Proof of Lemma 1

Using the law of iterated expectations, we have

$$\mathsf{E}\left[d(Z, \hat{z}(V,W))\right] = \mathsf{E}_V\left[\mathsf{E}\left[d(Z, \hat{z}(V,W))|V\right]\right]. \tag{15}$$

Now, for each $v \in \mathcal{V}$,

$$\mathsf{E}\left[d(Z, \hat{z}(V, W))|V = v\right] = \sum_{z \in \mathcal{Z}, w \in \mathcal{W}} p(z|v) p(w|v) d(z, \hat{z}(v, w))$$

$$= \sum_{w \in \mathcal{W}} p(w|v) \sum_{z \in \mathcal{Z}} p(z|v) d(z, \hat{z}(v, w))$$

$$\geq \min_{w \in \mathcal{W}} \sum_{z \in \mathcal{Z}} p(z|v) d(z, \hat{z}(v, w)) \qquad (16)$$

$$= \sum_{z \in \mathcal{Z}} p(z|v) d(z, \hat{z}(w^*(v))),$$

where $w^*(v)$ attains the minimum in (16) for a given $v$. Define $\hat{z}^*(v) = \hat{z}(w^*(v))$. Then (15) becomes

$$\mathsf{E}\left[d(Z, \hat{z}(V, W))\right] = \mathsf{E}_V\left[\mathsf{E}\left[d(Z, \hat{z}(V, W))|V\right]\right]$$

$$\geq \mathsf{E}_V\left[\sum_{z \in \mathcal{Z}} p(z|v) d(z, \hat{z}^*(v))\right]$$

$$= \mathsf{E}\left[d(Z, \hat{z}^*(V))\right]$$

which completes the proof.

# APPENDIX B

## PROOF OF THEOREM 2

Before proving the Theorem 2, we summarize a few useful properties of $C_{\text{SC}}(D)$ in Lemma 2. In [31], they also discussed similar properties of the capacity–distortion function for the case in which the channel state information is not available.

*Lemma 2:* The capacity–distortion function $C_{\text{SC}}(D)$ in Theorem 2 has the following properties:

(1) $C_{\text{SC}}(D)$ is a nondecreasing concave function of $D$ for all $D \geq D^*$.

(2) $C_{\text{SC}}(D)$ is a continuous function of $D$ for all $D > D^*$.

(3) $C_{\text{SC}}(D^*) = 0$ if $D^* \neq 0$ and $C_{\text{SC}}(D^*) \geq 0$ if $D^* = 0$.

The monotonicity is trivial. The concavity can be shown by using the standard time sharing argument. The continuity is a direct consequence of the concavity. The last property follows from Section IV. With these properties in hand, let us prove Theorem 2.

## A. Proof of Achievability

We use $b$ transmission blocks, each consisting of $n$ symbols. The encoder uses rate-splitting technique, whereby in block $j$, it appropriately allocates its rate between transmitting independent information and a description of the state sequence $S^n(j-1)$ in block $j-1$.

**Codebook generation.** Fix a conditional pmf $p(x)p(u|x,s)$ and function $\hat{s}(u,x,y)$ that attain $C_{\text{SC}}(D/(1+\epsilon))$, where $D$ is the desired distortion, and let $p(u|x) = \sum_s p(s)p(u|x,s)$. For each $j \in [1:b]$, randomly and independently generate $2^{n(R+R_S)}$ sequences $x^n(m_j, l_{j-1})$, $m_j \in [1:2^{nR}], l_{j-1} \in [1:2^{nR_S}]$, each according to $\prod_{i=1}^n p_X(x_i)$. For each $m_j \in [1:2^{nR}], l_{j-1} \in [1:2^{nR_S}]$, randomly and conditionally independently generate $2^{n\tilde{R}_S}$ sequences $u^n(k_j|m_j, l_{j-1})$, $k_j \in [1:2^{n\tilde{R}_S}]$, each according to $\prod_{i=1}^n p_{U|X}(u_i|x_i(m_j, l_{j-1}))$. Partition the set of indices $k_j \in [1:2^{n\tilde{R}_S}]$ into equal-size bins $\mathcal{B}(l_j) = [(l_j-1)2^{n(\tilde{R}_S - R_S)} + 1 : l_j 2^{n(\tilde{R}_S - R_S)}]$, $l_j \in [1:2^{nR_S}]$. This defines the codebook

$$\mathcal{C}_j = \{(x^n(m_j, l_{j-1}), u^n(k_j|m_j, l_{j-1})): m_j \in [1:2^{nR}], l_{j-1} \in [1:2^{nR_S}], k_j \in [1:2^{n\tilde{R}_S}]\}, \quad j \in [1:b].$$

The codebook is revealed to the both encoder and the decoder.

**Encoding.** By convention, let $l_0 = 1$. At the end of block $j$, the encoder finds an index $k_j$ such that

$$(s^n(j), u^n(k_j|m_j, l_{j-1}), x^n(m_j, l_{j-1})) \in \mathcal{T}_{\epsilon'}^{(n)}.$$

If there is more than one such index, it selects one of them uniformly at random. If there is no such index, it selects an index from $[1:2^{n\tilde{R}_S}]$ uniformly at random. In block $j+1$ the encoder transmits $x^n(m_{j+1}, l_j)$, where $m_{j+1}$ is the new message index to be sent in block $j+1$ and $l_j$ is the bin index of $k_j$.

**Decoding.** Let $\epsilon > \epsilon'$. At the end of block $j+1$, the decoder finds the unique index $\hat{m}_{j+1}, \hat{l}_j$ such that $(x^n(\hat{m}_{j+1}, \hat{l}_j), y^n(j+1)) \in \mathcal{T}_\epsilon^{(n)}$. The decoder thus decodes the message index $\hat{m}_{j+1}$ in block $j+1$. It then finds the unique index $\hat{k}_j \in \mathcal{B}(\hat{l}_j)$ such that $(u^n(\hat{k}_j|\hat{m}_j, \hat{l}_{j-1}), x^n(\hat{m}_j, \hat{l}_{j-1}), y^n(j)) \in \mathcal{T}_\epsilon^{(n)}$. Finally it computes the reconstruction sequence as $\hat{s}_i(j) = \hat{s}(u_i(\hat{k}_j|\hat{m}_j, \hat{l}_{j-1}), x_i(\hat{m}_j, \hat{l}_{j-1}), y_i(j))$ for $i \in [1:n]$.

Following the analysis of minimum distortion in Section II, it can be readily shown that the scheme can achieve any rate up to the capacity–distortion function given in Theorem 2.

## B. Proof of the Converse

We need to show that given any sequence of $(2^{nR}, n)$ code with $\lim_{n \to \infty} P_e^{(n)} = 0$ and $\mathsf{E}(d(S^n, \hat{S}^n)) \leq D$, we must have $R \leq C_{\text{SC}}(D)$. We identify the auxiliary random variables $U_i := (M, S^{i-1}, Y_{i+1}^n)$, $i \in [1:n]$ with $S_0 = Y_{n+1} = \emptyset$. Note that, as desired, $U_i \to (X_i, S_i) \to Y_i$ form a Markov chain for $i \in [1:n]$. Consider

$$\begin{aligned}
nR &= H(M) \\
&\stackrel{(a)}{\leq} I(M; Y^n) + n\epsilon_n \\
&= \sum_{i=1}^n I(M; Y_i | Y_{i+1}^n) + n\epsilon_n \\
&\leq \sum_{i=1}^n I(M, Y_{i+1}^n; Y_i) + n\epsilon_n \\
&= \sum_{i=1}^n I(M, Y_{i+1}^n, S^{i-1}; Y_i) - \sum_{i=1}^n I(S^{i-1}; Y_i | M, Y_{i+1}^n) + n\epsilon_n \\
&\stackrel{(b)}{=} \sum_{i=1}^n I(M, Y_{i+1}^n, S^{i-1}, X_i; Y_i) - \sum_{i=1}^n I(S^{i-1}; Y_i | M, Y_{i+1}^n) + n\epsilon_n \\
&\stackrel{(c)}{=} \sum_{i=1}^n I(M, Y_{i+1}^n, S^{i-1}, X_i; Y_i) - \sum_{i=1}^n I(Y_{i+1}^n; S_i | M, S^{i-1}) + n\epsilon_n \\
&\stackrel{(b)}{=} \sum_{i=1}^n I(M, Y_{i+1}^n, S^{i-1}, X_i; Y_i) - \sum_{i=1}^n I(Y_{i+1}^n; S_i | M, S^{i-1}, X_i) + n\epsilon_n \\
&\stackrel{(d)}{=} \sum_{i=1}^n I(M, Y_{i+1}^n, S^{i-1}, X_i; Y_i) - \sum_{i=1}^n I(M, S^{i-1}, X_i, Y_{i+1}^n; S_i) + n\epsilon_n \\
&= \sum_{i=1}^n I(U_i, X_i; Y_i) - \sum_{i=1}^n I(U_i, X_i; S_i) + n\epsilon_n \quad (17)
\end{aligned}$$

where $(a)$ follows by Fano's inequality [8, Theorem 7.7.1], which states that $H(M|Y^n) \leq n\epsilon_n$ for some $\epsilon_n \to 0$ as $n \to \infty$ for any code satisfying $\lim_{n \to \infty} P_e^{(n)} = 0$, $(b)$ follows since $X_i$ is a function of $(M, S^{i-1})$, $(c)$ follows by the Csiszár sum identity [9], [13], [11, Sec. 2.3], and $(d)$ follows since $(M, S^{i-1}, X_i)$ is independent of $S_i$. So now we have

$$\begin{aligned}
R &\leq \frac{1}{n}\sum_{i=1}^n I(U_i, X_i; Y_i) - \sum_{i=1}^n I(U_i, X_i; S_i) + n\epsilon_n \\
&\stackrel{(a)}{\leq} \frac{1}{n}\sum_{i=1}^n C_{\text{SC}}(\mathsf{E}(d(S_i, \hat{s}_i(U_i, X_i, Y_i)))) + n\epsilon_n \\
&\stackrel{(b)}{\leq} C_{\text{SC}}\Big(\frac{1}{n}\sum_{i=1}^n \mathsf{E}(d(S_i, \hat{s}_i(U_i, X_i, Y_i)))\Big) + n\epsilon_n \\
&\stackrel{(c)}{\leq} C_{\text{SC}}(D), \quad (18)
\end{aligned}$$

where $(a)$ follows from the definition of the capacity–distortion function, $(b)$ follows by the concavity of $C_{\text{SC}}(D)$ (see Property 1 in Lemma 2), and $(c)$ follows from Lemmas 1 and 2. This completes the proof of Theorem 2.